\documentclass[preprint2]{aastex}
\usepackage{psfig,natbib}
\shortauthors{Bower et al.}
\shorttitle{Linear Polarization of Sgr~A*}
\begin{document}

\newcommand\degd{\ifmmode^{\circ}\!\!\!.\,\else$^{\circ}\!\!\!.\,$\fi}
\newcommand{\etal}{{\it et al.\ }}
\newcommand{\uv}{(u,v)}
\newcommand{\rdm}{{\rm\ rad\ m^{-2}}}

\title{BIMA Observations of Linear Polarization in Sagittarius A* at 112 GHz}

\author{Geoffrey C. Bower\altaffilmark{1,2}, 
Melvyn C.H. Wright\altaffilmark{2},
Heino Falcke\altaffilmark{3}, 
Donald C. Backer\altaffilmark{2} }

\altaffiltext{1}{National Radio Astronomy Observatory, P.O. Box O, 1003 
Lopezville, Socorro, NM 87801; gbower@nrao.edu} 
\altaffiltext{2}{Astronomy Department \& Radio Astronomy Laboratory, 
University of California, Berkeley, CA 94720; gbower,mwright,dbacker@astro.berkeley.edu}
\altaffiltext{3}{Max Planck Institut f\"{u}r Radioastronomie, Auf dem 
H\"{u}gel 69, D 53121 Bonn Germany; hfalcke@mpifr-bonn.mpg.de} 

\begin{abstract}

We report here BIMA array observations of linear polarization in
Sagittarius A*, the compact radio source in the Galactic Center.
These observations had a resolution of $20^{\prime\prime} \times 5^{\prime\prime}$
oriented North-South.
We do not detect linear or circular polarization at a level of 1.8\% 
(1-$\sigma$) at 112 GHz.
This puts a new constraint on models that show a sharp change in
linear polarization fraction at 100 GHz.

\end{abstract}


\section{Introduction}

The evidence that Sagittarius A* is associated with a supermassive
black hole in the Galactic Center has been reviewed recently
in \citet{1998ApJ...494L.181M} and \citet{1999ApJ...524..805B}. 
However, significant details of the accretion and emission region are not
understood.  In particular, it is not known whether the emission
originates in an inflow or an outflow and whether the emission
mechanism is synchrotron or gyrosynchrotron 
\citep[e.g.,][]{1994ApJ...426..577M,2000ApJ...541..234O,2000A&A...362..113F}.

Polarization is an important diagnostic of Sgr~A*.  The source
exhibits no linear polarization at frequencies less than 86 GHz
\citep{1999ApJ...521..582B,1999ApJ...527..851B}.
Interstellar depolarization
has been eliminated as an effect by this high frequency result and by
4.8 and 8.4 GHz spectropolarimetry.  Recently, 
\citet{2000ApJ...534L.173A}
have claimed detection of high levels of linear polarization
in Sgr~A* at millimeter and submillimeter wavelengths.  Their
JCMT/SCUBA observations at 150 GHz indicate $p=12^{+9}_{-4}\%$ and
increasing values at higher frequencies.  The low resolution
of these observations ($33.5^{\prime\prime}$) required substantial
corrections to the observed total intensity and polarized
fluxes due to dust and free-free emission in order to determine
the polarization of Sgr~A*.  The implied sharp rise in polarization
has significant implications for models of the emission
region 
\citep{2000ApJ...545..842Q,2000ApJ...538..L121,2000ApJ...545L.117M}.

We have also pursued circular polarization observations of Sgr~A*
\citep{1999ApJ...523L..29B,2001ApJprepbower}.
These results indicate
that the circular polarization is variable and that it
exceeds the linear polarization
at frequencies as high as 15 GHz and possibly higher.
Any complete model of the emission region must account for both
linear and circular polarization.

We present here BIMA array
observations of linear and circular polarization in Sgr~A* at 112 and 115 GHz.
These interferometric observations have a resolution of
$20^{\prime\prime} \times 5^{\prime\prime}$, which is an order
of magnitude less than those of Aitken etal.  We detect no
linear or circular polarization at a level of 1\%.

\section{Observations and Results}

We used the BIMA array in its C array configuration
on 13 May 2000 to observe Sgr~A* and on 12 May and 6 June 2000 
to observe the 
polarization calibrators 3C 273 and 3C 279.  Observations were made in two
sidebands with central frequencies of 111.8 and 115.3 GHz
and bandwidths of 800 MHz each.  

We observed in the polarization switching mode
previously described in \citet{1999ApJ...527..851B}.  A single antenna was kept
in linear polarization while other antennas switched between
right (RCP) and left (LCP) circular polarization.  Self-calibration
of parallel-hand visibilities was performed with respect to
the reference antenna in linear polarization.  The compact
and highly polarized source 3C 279 was used as a polarization 
leakage calibrator on 12 May and 6 June.  Observations were 3.5 and 4.5
h in duration on the two dates, respectively.  We plot 
the lower sideband 
polarization leakage solutions in Figure~\ref{fig:leak}.  The
rms difference between these solutions is 1.3\%.
This indicates a maximum 1-$\sigma$ systematic error of 
0.4\% for polarization results.  Previous measurements of
linear polarization with BIMA have achieved similar limits
\citep[e.g.,][]{1999ApJ...527..851B,1998ApJ...502L..75R}.  

These leakage solutions were applied to
the phase-self-calibrated data for Sgr~A* and other calibrators. 
The calibrated visibilities were imaged and deconvolved 
in all four Stokes parameters.
Baseline lengths ranged from 10 to 80 m.  This produced a resolution of 
$20^{\prime\prime} \times 5^{\prime\prime}$ in position angle $-3.4^\circ$
for Sgr~A*.  
We tabulate the total intensity and polarized fluxes in the images
Table~\ref{tab:fluxes}.  These fluxes were measured through the
fitting of Gaussians at the phase-center of the images.  
Errors in linear and circular polarization
are the 1-$\sigma$ statistical error.  

Differences in the polarization
between epochs for some sources may be due to real source variability.
Compact, flat-spectrum sources are known to show significant
changes in their total and polarized flux density on timescales
less than 2 months \citep{1999AAS...195.8902M}.

The results at 112 GHz are clearly more reliable than those at 115 GHz.
For Sgr~A*, the higher frequency flux is $\sim 0.5$ times the flux at
112 GHz and the polarization measurement error is higher by a factor of 
$\sim 4$.  Similar effects are seen for the other sources.  
This is the result of increased atmospheric opacity from the 118 GHz
O$_2$ line.   System temperatures at 115 GHz were roughly
twice those at 112 GHz, as predicted for model atmospheres 
\citep[e.g.,][]{1989IJIMW..10..631L}.
The increased system temperature leads to lower SNR in self-calibration
of the antenna phases, and thus decorrelation.
The decorrelation is enhanced for Sgr~A*
due its low declination.  Therefore, while the polarization results at
the two frequencies are consistent, the more accurate results are at
112 GHz.

Decorrelation is unlikely to play a large
effect at 112 GHz.  The short baseline
flux that we measure for Sgr~A* at 112 GHz is roughly equal to the
peak flux in the JCMT maps at 150 GHz with $33.5^{\prime\prime}$
resolution.  Furthermore, analysis of J1733-130 and the other calibrators
indicates that there is less than
25\% decorrelation on the longest baselines with no dependence 
on source declination.  

Confusion of thermal emission from Sgr~A West may increase
the measured total intensity of Sgr~A*.  This confusion is likely 
to be small with respect to the total intensity of Sgr~A*.  
This is in part because the good East-West resolution separates
Sgr~A* from the North-South arm of the minispiral.  There is
some component of emission in the East-West arm, or bar, of the minispiral,
however.  We can estimate a maximum confusion flux from the 
well-sampled 3.6 cm VLA continuum image of Sgr~A West of
\citet{1993ApJS...86..133R}.
Convolving this image 
with a $20^{\prime\prime} \times 5^{\prime\prime}$ beam we find a total
flux at the map center of $ 5.4$ Jy.  The actual flux of Sgr~A*
in this map is 0.7 Jy, implying an excess flux of $ 4.7$ Jy.  Extrapolating
with a power law index appropriate to free-free emission 
$(S\propto \nu^{-0.1})$, we expect 3.6 Jy of
additional flux in the central beam.  Clearly, much of this flux
is absent in our image.  This is in part a result of the
less complete $\uv$-coverage of our observations, especially on 
short-baselines.  Extended flux is not usually preserved in images
made from incompletely-sampled visibility data 
\citep[e.g.,][]{1999sira.conf..151C}.  This flux can be seen in the
$\uv$-distance plot on baselines shorter than 10 $k\lambda$ 
(Figure~\ref{fig:uvdamp}). 

A more direct comparison comes from an analysis of the visibility
amplitudes as a function
of $\uv$-distance and a comparision with higher resolution results at 
lower frequencies 
\citep{1999ApJ...527..851B,1999ApJ...523L..29B,2001ApJprepbower}.
These show that flux in the visibility range of 20 to 35 $k\lambda$ is
an excellent estimator of the total flux over a broad range of 
frequencies.
In VLA A-array observations of Sgr~A* at 8.4 GHz,
the mean flux between 20 and 35 $k\lambda$ is $0.78 \pm 0.13$ Jy,  
which is very close to the true flux of $0.75 \pm 0.02$ Jy 
determined on baselines longer than 100 $k\lambda$.  
At 43 GHz, VLA A-array fluxes in the same two $\uv$ distance
ranges are $0.92\pm 0.16$ Jy and $0.99 \pm 0.02$ Jy, respectively.
And at 86 GHz, BIMA fluxes are $2.27 \pm 0.47$ Jy and $2.40 \pm 0.02$ Jy, 
respectively.  This effect is due to the averaging of flux from arcsecond-scale
structure that beats with the flux from Sgr~A*.  Apparently, and as expected,
it is independent of frequency, specific $\uv$-coverage and the flux of Sgr A*.

For our 112 GHz data 
the flux density is $0.85 \pm 0.04$ Jy in the range 20 to 35 $k\lambda$.
The consistency with the lower frequency data sets is shown in 
Figure~\ref{fig:uvdamp}. 
This implies a free-free contribution to the flux measured
at the position of Sgr~A* to be $\sim 0.5$ Jy.  The 1-$\sigma$
upper limit for the polarization of Sgr~A* is 1.8\%.

\section{Discussion}

These results indicate that the linear polarization of Sgr~A*
at 112 GHz is not detected at a level of 1.8\%.  This is consistent
with previously-made lower frequency observations 
\citep{1999ApJ...521..582B,1999ApJ...527..851B}.
We can reject the hypothesis of 
constant linear polarization fraction at a weighted mean of 3.5\%
between 112 and 150 GHz:  our measurement combined
with the JCMT/SCUBA result gives $\chi^2_\nu=5.4$.
Thus, a very steep rise in linear polarization fraction is necessary to 
reconcile the two measurements.
As both \citet{2000ApJ...545..842Q} and \citet{2000ApJ...538..L121}
have shown, very specific conditions are necessary in the
source and in the accretion environment to produce such a steep rise.

Variability of the linear polarization signal is a potential explanation
for the discrepancy.  
\citet{2001ApJ...547L..29Z} have recently demonstrated 106-day periodic 
variations in the total intensity of Sgr~A* at 15 and 22 GHz.  The 
strength of the variation does increase with frequency, suggesting that
there may be substantial variation at frequencies above 100 GHz.  
However, the number of epochs of polarization
monitoring indicate that this is an unlikely cause.  We observed
in total four separate epochs at 86 and 112 GHz without detection.
The flux varied by  a factor of two between these
observations with no apparent change in the linear polarization.
Such variations in the total flux have been observed previously
at high frequencies \citep{1993ApJ...417..560W,1999cpg..conf..105T}.
Additionally, \citet{2000ApJ...534L.173A} claim detection at
different frequencies in two widely separated
epochs.  Interestingly, these two epochs are roughly at the same phase of the
106-day period.  If variability were confirmed as the source of the
discrepancy, then the relationship between total intensity and polarized
intensity variations would be a very important diagnostic for Sgr~A*.

We also present measurements of the circular polarization of these
sources.  Circular polarization is not detected in any source.
For Sgr~A*, the rms uncertainty in the polarization is 
$1.7\%$.  The high degree of order in the magnetic field
necessary to produce linear polarization fractions greater than 10\%
will also lead to a higher fractional circular polarization in
a synchrotron or gyrosynchrotron source.  
Thus, the absence of circular polarization also suggests that the
linear polarization limit is robust.  The circular polarization 
is known to be variable with an inverted spectrum
at frequencies as high as 15 GHz
\citep{2001ApJprepbower}.

An absence of linear polarization at high frequencies has several
possible explanations.  One, a dense accretion region as expected by
both jet and ADAF models will likely depolarize a linearly polarized
signal.  The low accretion rate  model of 
\citet{2000ApJ...545..842Q} is not required.  Two, the emission region  may have a strongly tangled 
magnetic field.  Three, the emission is not synchrotron or there is 
a large number of low energy electrons in the source that depolarize
the emission.

Higher frequency interferometric observations are crucial to resolve the
apparent disagreement between these results and those of 
\citet{2000ApJ...534L.173A}.  Beyond that, connection of linear and 
circular polarization properties with each other and with total
intensity variability stands as an important observational 
and theoretical goal for our understanding of Sgr~A*.

\acknowledgements

The BIMA array is operated by the
Berkeley-Illinois-Maryland Association under funding from the National
Science Foundation.


\begin{deluxetable}{llrrrrrrrrr}
\tablecaption{Polarized Flux of All Sources \label{tab:fluxes}}
\tablehead{ \colhead{Source} &\colhead{Date} &  \colhead{$\nu$} & 
\colhead{$I$} & \colhead{$\sigma_I$} & \colhead{$p$} & \colhead{$\sigma_p$} &
\colhead{$\chi$} & \colhead{$\sigma_\chi$} & \colhead{$V$}  & \colhead{$\sigma_V$} \\
                           &                 &\colhead{(GHz)}   &
\colhead{(Jy)} & \colhead{(Jy)}      & \colhead{(\%)}& \colhead{(\%)}       &
\colhead{(deg)}  & \colhead{(deg)}        & \colhead{(\%)}  & \colhead{(\%)} }
\startdata
3C 273     & May12 & 112 &    6.665 &   0.008 &    3.18 &    0.12 &  -44.9 &    1.1 &    0.33 & 0.12\\
           &       & 115 &    5.316 &   0.010 &    3.25 &    0.19 &  -46.4 &    1.7 &    0.24 & 0.19\\
           & Jun06 & 112 &    6.737 &   0.011 &    2.69 &    0.17 &  -26.8 &    1.8 &   -0.38 & 0.17\\
           &       & 115 &    5.359 &   0.015 &    3.15 &    0.28 &  -30.2 &    2.6 &   -0.42 & 0.28\\
3C 279     & May12 & 112 &   19.080 &   0.007 &   11.33 &    0.03 &   61.3 &    0.1 &    0.07 & 0.03\\
           &       & 115 &   15.630 &   0.009 &   11.40 &    0.06 &   62.1 &    0.1 &    0.08 & 0.06\\
           & Jun06 & 112 &   15.380 &   0.006 &   14.12 &    0.04 &   58.8 &    0.1 &    0.40 & 0.40\\
           &       & 115 &   12.520 &   0.009 &   14.23 &    0.07 &   58.4 &    0.1 &    0.40 & 0.07\\
3C 454.3   & May13 & 112 &    5.390 &   0.014 &    1.97 &    0.26 &  -80.2 &    3.8 &   -0.15 & 0.26\\
           &       & 115 &    4.341 &   0.019 &    0.96 &    0.45 &  -74.6 &   13.3 &   -0.85 & 0.45\\
           & Jun06 & 112 &    4.733 &   0.014 &    1.37 &    0.30 &  -65.9 &    6.2 &   -0.26 & 0.30\\
           &       & 115 &    3.921 &   0.019 &    1.50 &    0.50 &  -89.1 &    9.5 &   -1.06 & 0.50\\
J1733-130  & May13 & 112 &    1.719 &   0.018 &    6.14 &    1.06 &   -6.3 &    4.9 &    1.36 & 1.06\\
           &       & 115 &    0.968 &   0.027 &    8.45 &    2.82 &  -15.4 &    9.6 &    4.57 & 2.82\\
           & Jun06 & 112 &    1.622 &   0.022 &    4.64 &    1.33 &   -6.1 &    8.2 &    0.96 & 1.33\\
           &       & 115 &    1.266 &   0.032 &    7.02 &    2.56 &   25.7 &   10.4 &   -6.62 & 2.56\\
Sgr A*     & May13 & 112 &    1.412 &   0.015 &    1.16 &    1.09 &   75.2 &   26.8 &    0.19 & 1.09\\
           &       & 115 &    0.735 &   0.030 &    2.79 &    4.13 &  -31.6 &   42.5 &    0.04 & 4.13\\
\enddata

\end{deluxetable}

\newpage

\plotone{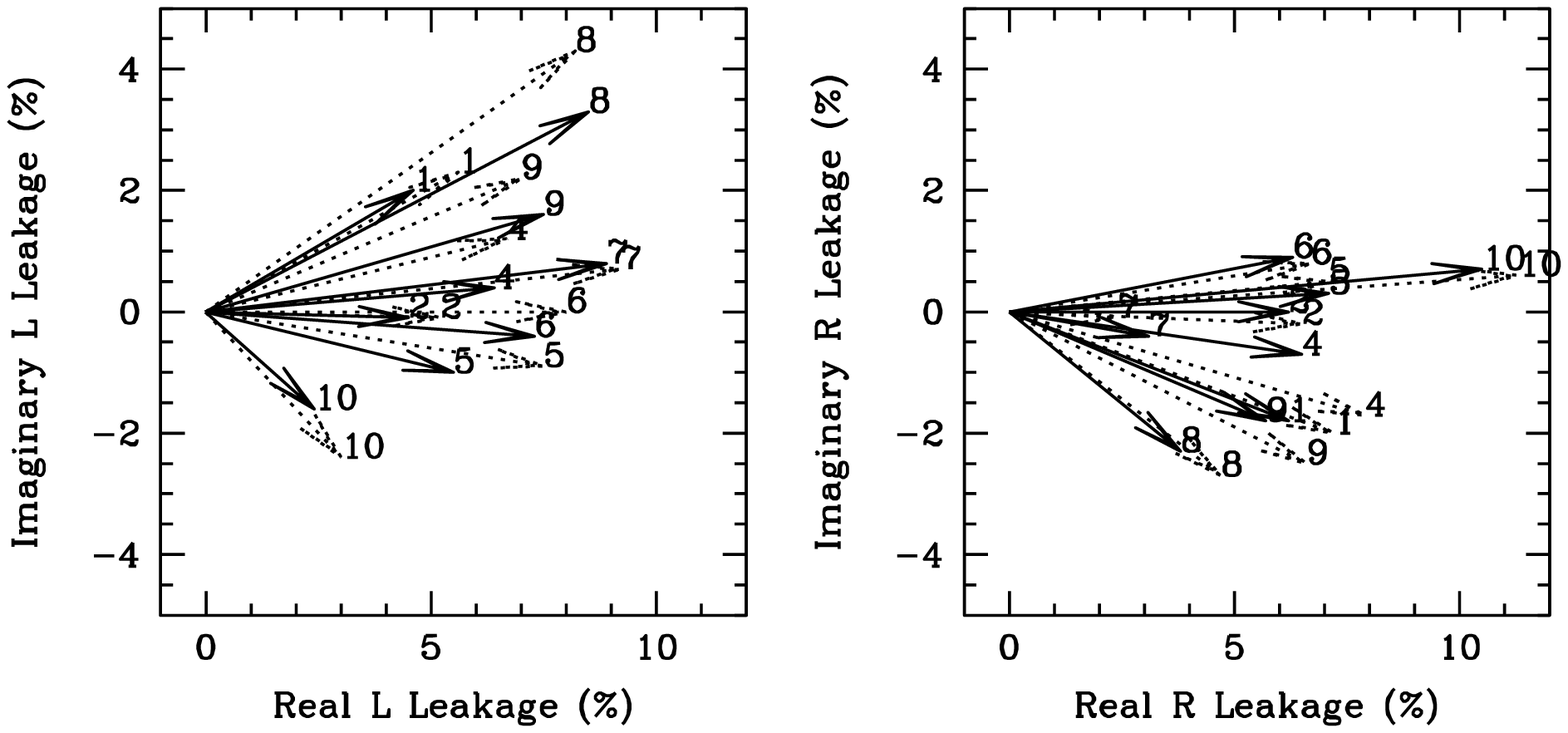}
\figcaption[leakcompare.ps]{The leakage terms for each antenna in the
BIMA array at 112 GHz determined by observations of 3C 279 on
12 May 2000 (solid line) and 6 June 2000 (dotted line).
\label{fig:leak}}

\plotone{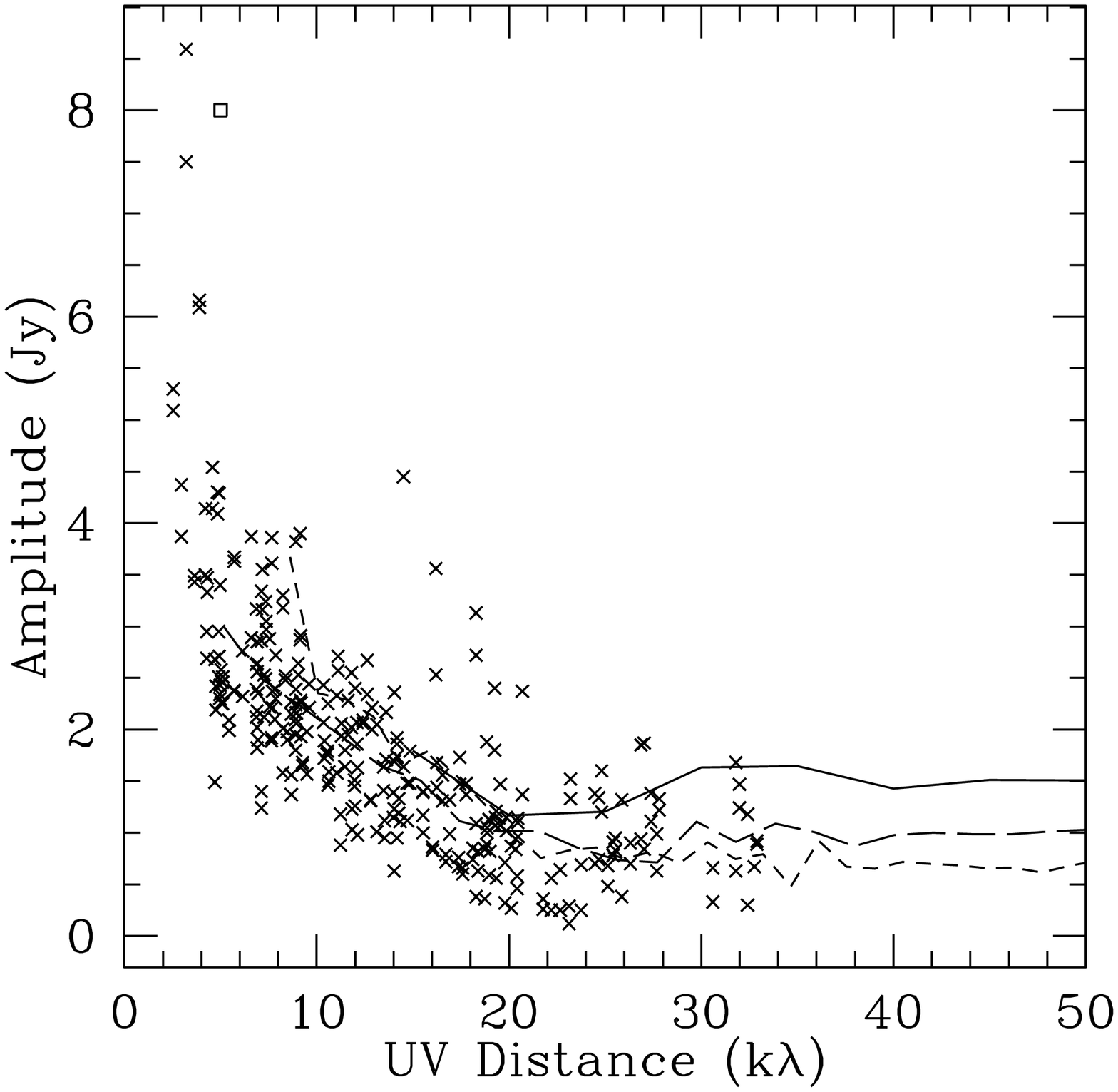}
\figcaption[uvdamp_new.ps]{The calibrated total intensity amplitude in Jy for
Sgr A* at 112 GHz as a function of projected baseline length.  The
short-dashed, long-dashed and solid lines 
indicate mean visibility amplitudes from 
previous 8.4, 43 and 86 GHz observations with high resolution, respectively.  
The measured fluxes were $0.75\pm 0.02$ Jy, $0.99 \pm 0.02$ Jy 
and $2.40\pm 0.02$ Jy, respectively.  Note that mean flux in the 20 to 35
$k\lambda$ range is a good estimator of the long baseline flux.
The 86 GHz results were plotted with 1 Jy subtracted.
\label{fig:uvdamp}}

\end{document}